\documentclass[english,twocolumn, aps]{revtex4-1}
\usepackage[T1]{fontenc}
\usepackage[latin9]{inputenc}
\usepackage{babel}
\usepackage{amsmath}
\usepackage{amssymb}
\usepackage{graphicx}
\usepackage{esint}
\usepackage[unicode=true,pdfusetitle,
 bookmarks=true,bookmarksnumbered=false,bookmarksopen=false,
 breaklinks=false,pdfborder={0 0 1},backref=false,colorlinks=false]
 {hyperref}
\usepackage{breakurl}
\begin{document}

\title{Thickness-dependent Kapitza resistance in multilayered graphene and
other two-dimensional crystals}

\author{Zhun-Yong Ong}

\email{ongzy@ihpc.a-star.edu.sg}

\affiliation{Institute of High Performance Computing, A{*}STAR, Singapore 138632,
Singapore}
\begin{abstract}
The Kapitza or thermal boundary resistance (TBR), which limits heat
dissipation from a thin film to its substrate, is a major factor in
the thermal management of ultrathin nanoelectronic devices and is
widely assumed to be a property of only the interface. However, data
from experiments and molecular dynamics simulations suggest that the
TBR between a multilayer two-dimensional (2D) crystal and its substrate
decreases with increasing film thickness. To explain this thickness
dependence, we generalize the recent theory for single-layer 2D crystals
by Z.-Y. Ong \emph{et al.} {[}Phys. Rev. B 94, 165427 (2016){]}, which
is derived from the theory by B. N. J. Persson \emph{et al.} {[}J.
Phys.: Condens. Matter 23, 045009 (2011){]}, and use it to evaluate
the TBR between bare $N$-layer graphene and SiO$_{2}$. Our calculations
reproduce quantitatively the TBR thickness dependence seen in experiments
and simulations as well as its asymptotic convergence, and predict
that the low-temperature TBR scales as $T^{-4}$ in few-layer graphene.
Analysis of the interfacial transmission coefficient spectrum shows
that the TBR reduction in few-layer graphene is due to the additional
contribution from higher flexural phonon branches. Our theory sheds
light on the role of flexural phonons in substrate-directed heat dissipation
and provides the framework for optimizing the thermal management of
multilayered 2D devices. 
\end{abstract}
\maketitle

\section{Introduction}

The Kapitza or thermal boundary resistance~\cite{Swartz:RMP89_Thermal}
(TBR) $R_{\text{K}}$ between a two-dimensional (2D) crystal (e.g.
graphene) and its substrate (e.g. SiO$_{2}$) determines the ratio
of the temperature difference at the interface ($\Delta T$) to the
rate of phonon-mediated heat transfer from the 2D crystal to the substrate
($Q$), i.e. $R_{\text{K}}=\Delta T/Q$, and thus plays a key role
in the control of heat dissipation from atomically thin nanoelectronic
devices~\cite{Freitag:NL09_Energy,Pop:NResearch10_Energy,Bae:NL10_Imaging,Bae:NComm13_Ballistic,Serov:JAP14_Theoretical,Li:NMTE15_PhononDynamics}.
In particular, the lattice thermal boundary conductance (TBC) $G$,
defined as $G=R_{\text{K}}^{-1}$, depends on the lattice properties
of the materials as well as their van der Waals (vdW) interaction.
Although variations in the Kapitza resistance have been attributed~\cite{Cai:NL10_Thermal,Persson:JPCM10_Heat,Persson:JPCM11_Phononic,Liu:AIDAdv15_ContactArea}
to differences in interfacial roughness and contact as well as to
remote phonon scattering~\cite{Ong:PRB13_Signature,Koh:NL16_Role},
there is increasing evidence from molecular dynamics (MD) simulations~\cite{Ni:APL13_Few,Liang:PRL14_Thermal,Liang:PRB14_FiniteSize,Chen:NL14_Strain,Ni:PRB14_Substrate,Alexeev:NL15_Kapitza}
and experiments~\cite{Mak:APL10_Measurement,Menges:PRL13_Thermal,Yang:PRL14_Graphitic,Yuan:Acta17_Interfacial}
that the TBR between an unencapsulated or \emph{bare} thin film and
its substrate has a strong thickness dependence for which we have
no satisfactory explanation especially in a layered crystal such as
graphene. It has been observed~\cite{Ni:APL13_Few} that the TBR
for few-layer graphene decreases significantly as the film thickness
(or the number of layers $N$) increases and asymptotically converges
to a fixed value when $N$ is large. This implies that single or few-layer
2D crystals are less efficient at dissipating heat to the substrate
and must be factored into the design of high-power nanoelectronic
devices.

However, elucidating the physics underlying this TBR thickness dependence
is a challenge given the awkwardness of describing \emph{single-sided}
cross-plane phonon transport in few-layer films using traditional
mismatch models~\cite{Swartz:RMP89_Thermal} when the two media are
dimensionally different~\cite{Correa:Nanotech17_Interface}. Although
MD simulations can reproduce the room-temperature TBR~\cite{Ni:APL13_Few,Ni:PRB14_Substrate},
they are computationally expensive, restricted to the classical regime
and constrained by size effects~\cite{Liang:PRB14_FiniteSize}, and
thus they lack direct insight into more fundamental phononic processes.
On the other hand, an analytical approach in which flexural phonons
are explicitly treated like in Refs.~\cite{Persson:JPCM11_Phononic,Ong:PRB16_Theory}
can shed more light on this thickness dependence by yielding new insights
into how the layered geometry and individual flexural phonon branches
affect interfacial heat dissipation, and thereby improve the control
of heat dissipation in layered films such as vdW heterostructures~\cite{Novoselov:Science16_2D}.
The continuum physics-based approach to investigating nanoscale interfacial
heat dissipation was first formulated by Persson and co-workers and
applied to the graphene/SiO$_{2}$ interface in Refs.~\cite{Persson:EPJE10_Heat,Persson:JPCM10_Heat,Persson:JPCM11_Phononic}
although their predicted TBR values are an order of magnitude too
large because of spurious singularities from the use of the weak-coupling
approximation. This treatment is refined in Ref.~\cite{Ong:PRB16_Theory}
where it is shown that the interfacial heat transport theory can be
reformulated into a more conventional Landauer form with a well-defined
transmission coefficient spectrum and that flexural phonon damping
is necessary to yield a \emph{finite} TBR for single-layer graphene.
In addition, the extended treatment enables us to consider the effects
of a \emph{superstrate}, such as a top SiO$_{2}$ layer, on the TBR, yielding
the insight that there is significant distinction between the TBR
of a bare 2D crystal and that of a SiO$_{2}$-encased sample. In Ref.~\cite{Ong:PRB16_Theory},
the calculated TBR values for bare and SiO$_{2}$-encased single-layer
graphene~\cite{Ong:PRB16_Theory} are in good agreement with experiments
and show that the incorporation of the SiO$_{2}$ superstrate results
in a substantial reduction of the TBR because of the enhanced cross-plane
phonon transmission.

In this work, we extend this continuum physics-based approach by generalizing
our recently developed theory of cross-plane substrate-directed heat
dissipation for \emph{bare} single-layer 2D crystals~\cite{Ong:PRB16_Theory}
to \emph{bare} $N$-layer 2D crystals and applying it to the graphene/SiO$_{2}$
interface. This generalization is essentially made by deriving the
Green's function for the flexural motion of the thermally active bottommost
sheet in an $N$-layer stack, as given in Eq.~(\ref{Eq:SheetNGreensFunction}),
to obtain the correct expression for the interfacial transmission
coefficient function {[}Eq.~(\ref{Eq:TransmissionFunction}){]} needed
for calculating the thermal boundary conductance. The additional layers
stacked on the bottommost sheet in the $N$-layer stack modify its
flexural response to the interfacial stress exerted by the substrate
and, as we shall show later, introduce new transmission channels associated
with the higher flexural phonon branches. 

Our objectives here are twofold: the first is to elaborate on the
theoretical techniques and concepts introduced in Ref.~\cite{Ong:PRB16_Theory}
to investigate cross-plane heat dissipation from a 2D crystal and
the second is to explain the thickness-dependent TBR seen in Refs.~\cite{Mak:APL10_Measurement,Menges:PRL13_Thermal,Ni:APL13_Few,Liang:PRL14_Thermal,Liang:PRB14_FiniteSize,Chen:NL14_Strain,Yang:PRL14_Graphitic,Ni:PRB14_Substrate,Alexeev:NL15_Kapitza,Yuan:Acta17_Interfacial}. We
thus limit the scope of our discussion to \emph{bare} 2D crystals,
\emph{i.e.}, multilayer structures without any superstrate such as
a top SiO$_{2}$ layer, in order to understand more precisely the
origin of the TBR difference between single and multilayer 2D crystals.
We do not discuss the Kapitza resistance for \emph{encased} multilayer
2D crystals where there is an additional semi-infinite oxide or metal
superstrate encapsulating the $N$-layer 2D crystal because the superstrate
also alters the flexural response of the bottommost sheet in contact
with the substrate and exerts a similar effect on the interfacial
transmission through the introduction of additional transmission channels~\cite{Ong:PRB16_Theory}.
Hence, the theory presented in this work does not apply to the configuration
in which there is a multilayer 2D crystal sandwiched between two solid
slabs like in Refs.~\cite{Zhao:JAP05_Lattice,Shen:PRB13_Heat}.

The organization of our paper is as follows. In Sec.~\ref{Sec:Theory},
we begin with a review of the basic theoretical framework established
in Ref.~\cite{Ong:PRB16_Theory} for describing flexural phonon-mediated
heat dissipation from a one-layer 2D crystal to its substrate. We
then show through the derivation of Eqs.~ (\ref{Eq:SheetNGreensFunction})
and (\ref{Eq:IdenticalSheetNGreensFunction}) how the theory can be
generalized for a bare $N$-layer 2D crystal or van der Waals heterostructure
in which the flexural property of the thermally active bottommost
layer is modified by the layers stacked on top of it. In Sec.~\ref{Sec:Results-and-discussion},
the theory is applied to the study of the TBR of the $N$-layer graphene/SiO$_{2}$
interface and reproduces the TBR reduction observed in thicker graphene
samples which we attribute to enhanced interfacial phonon transmission.
Our analysis of the transmission coefficient spectra shows that the
enhanced transmission and the reduced TBR are due to the more efficient
transmission of the higher-frequency flexural phonon branches present
in $N$-layer graphene. We also show that the TBR converges asymptotically
as expected when $N\rightarrow\infty$, giving us the projected TBR
of the graphite/SiO$_{2}$ interface. The theoretical concepts and
techniques presented in this work are expected to be useful for the
analysis and interpretation of the TBR of multilayered structures
such as few-layer graphene and van der Waals heterostructures~\cite{Geim:Nature2013_VdW}.

\section{Theory\label{Sec:Theory}}

\subsection{Basic framework for  one-layer 2D crystals}

To describe mathematically heat dissipation from a multilayer 2D crystal,
we build on the framework introduced in Ref.~\cite{Ong:PRB16_Theory}.
The TBC between a single-layer 2D crystal and its substrate {[}see
Fig.~\ref{Fig:ModelSchematics}(a){]} is expressed in the Landauer
form as~\cite{Ong:PRB16_Theory}
\begin{equation}
G(T)=\frac{1}{(2\pi)^{3}}\int d^{2}q\int_{0}^{\infty}d\omega\hbar\omega\frac{\partial f(\omega,T)}{\partial T}\Xi(\mathbf{q},\omega)\ ,\label{Eq:ThermalBoundaryConductance}
\end{equation}
where $f(\omega,T)=[\exp(\hbar\omega/k_{B}T)-1]^{-1}$ is the Bose-Einstein
distribution function at frequency $\omega$ and temperature $T$.
The spectrum of the interfacial \emph{transmission coefficient} function
$\Xi(\mathbf{q},\omega)$, where 
\begin{equation}
\Xi(\mathbf{q},\omega)=\frac{4K^{2}\text{Im}D_{\text{sub}}(\mathbf{q},\omega)\text{Im}D_{\text{2D}}(\mathbf{q},\omega)}{|1-K[D_{\text{sub}}(\mathbf{q},\omega)+D_{\text{2D}}(\mathbf{q},\omega)]|^{2}}\label{Eq:TransmissionFunction}
\end{equation}
and $\mathbf{q}$ is the transverse wave vector, provides an intuitive
depiction of the phononic contributions to interfacial transmission
and satisfies the constraint $0<\Xi(\mathbf{q},\omega)\leq1$. In
Eq.~(\ref{Eq:TransmissionFunction}), $K$ is the spring constant
per unit area at the interface between the 2D crystal and the substrate
while $D_{\text{sub}}(\mathbf{q},\omega)$ and $D_{\text{2D}}(\mathbf{q},\omega)$
are the retarded Green's functions for the substrate surface and the
single-layer 2D crystal, respectively, that describe their flexural
or out-of-plane displacement response to the interfacial stress $\sigma_{\text{int}}(\mathbf{q},\omega)$.
 Assuming an isotropic elastic solid substrate, we have~\cite{Ong:PRB16_Theory,Persson:JCP01_Theory,Persson:JPCM11_Phononic}
\begin{equation}
D_{\text{\text{sub}}}(\mathbf{q},\omega)=\frac{i}{\rho_{\text{sub}}c_{T}^{2}}\frac{p_{L}(\mathbf{q},\omega)}{S(\mathbf{q},\omega)}\left(\frac{\omega}{c_{T}}\right)^{2}\label{Eq:SubstrateGreensFunction}
\end{equation}
where $S(\mathbf{q},\omega)=[(\omega/c_{T})^{2}-2q^{2}]^{2}+4q^{2}p_{T}p_{L}$,
$p_{L}(\mathbf{q},\omega)=\lim_{\eta\rightarrow0^{+}}[(\omega/c_{L})^{2}-q^{2}+i\eta]^{1/2}$,
$p_{T}(\mathbf{q},\omega)=\lim_{\eta\rightarrow0^{+}}[(\omega/c_{T})^{2}-q^{2}+i\eta]^{1/2}$,
and $c_{L}$, $c_{T}$ and $\rho_{\text{sub}}$ are its longitudinal
and transverse velocities, and its mass density per unit volume, respectively. 

If the 2D crystal is bare, \emph{i.e.}, it has no superstrate, then
$D_{\text{2D}}(\mathbf{q},\omega)$ in Eq.~(\ref{Eq:TransmissionFunction})
can be written as~\cite{Amorim:PRB13_Flexural,Ong:PRB16_Theory}
\begin{equation}
D_{\text{2D}}(\mathbf{q},\omega)=[\rho\omega^{2}+i\rho\gamma\omega-\kappa q^{4}]^{-1}\ ,\label{Eq:BareGreensFunc}
\end{equation}
where $\rho$, $\kappa$ and $\gamma$ are the mass density per unit
area, the bending rigidity and the frequency-dependent flexural damping
function of the 2D crystal, respectively. On the other hand, if there
is a superstrate such as an oxide layer on top of the 2D crystal (\emph{i.e.}
the 2D crystal is encased), then we have $D_{\text{2D}}(\mathbf{q},\omega)=\overline{D}_{\text{2D}}(\mathbf{q},\omega)$,
where
\begin{equation}
\overline{D}_{\text{2D}}(\mathbf{q},\omega)=[\rho\omega^{2}+i\rho\gamma\omega-\kappa q^{4}-P(\mathbf{q},\omega)]^{-1}\label{Eq:EncasedGreensFunc}
\end{equation}
is the retarded Green's function of the encased 2D sheet, $P(\mathbf{q},\omega)=g_{\text{top}}[1-g_{\text{top}}D_{\text{\text{top}}}(\mathbf{q},\omega)]^{-1}$
is the `self-energy' contribution from coupling to the superstrate,
and $g_{\text{top}}$ is the spring constant per unit area at the
top interface. The function $D_{\text{\text{top}}}(\mathbf{q},\omega)$
represents the retarded Green's function for the bottom surface of
the superstrate \emph{uncoupled to the thermally active 2D crystal
sheet immediately beneath it}. For a superstrate that is a semi-infinite
isotropic elastic solid~\cite{Ong:PRB16_Theory}, $D_{\text{\text{top}}}(\mathbf{q},\omega)$
has the same functional form as Eq.~(\ref{Eq:SubstrateGreensFunction})
although the functional form can be different for other superstrates.

\subsection{Generalization to two- and $N$-layer 2D crystals}

\begin{figure}
\includegraphics[width=8.5cm]{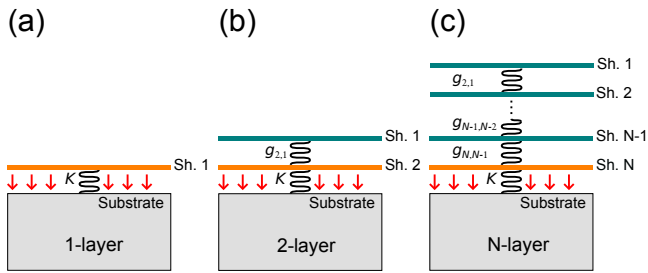}

\caption{Schematic of the substrate-supported (a) one-layer, (a) two-layer
and (c) $N$-layer 2D crystal. Adjacent sheets are connected to each
other by a harmonic force (represented by the springs). In our convention,
the individual sheets in the ``superstrate'' are numbered from $1$
to $N-1$ with sheet $1$ (`Sh. $1$') being the topmost sheet and
sheet $N$ (`Sh. $N$') being the bottommost sheet. Direct interfacial
heat transfer occurs only between the thermally active bottom sheet
and the substrate. }
\label{Fig:ModelSchematics}
\end{figure}

To generalize the TBC model in Ref.~\cite{Ong:PRB16_Theory}, we
first discuss the case for the two-layer crystal {[}see Fig.~\ref{Fig:ModelSchematics}(b){]},
which can be a homogeneous material like bilayer graphene or a composite
like a graphene/MoS$_{2}$ vdW heterostructure~\cite{Liu:RSCAdv15_Thermal}.
We do not assume that the two layers are identical for the sake of
generality. The TBC calculation using Eqs.~(\ref{Eq:ThermalBoundaryConductance})
and (\ref{Eq:EncasedGreensFunc}) requires us to determine $\overline{D}_{\text{2D}}(\mathbf{q},\omega)$
and $D_{\text{top}}(\mathbf{q},\omega)$. In our approach, we treat
the two-layer system as an encased one-layer 2D crystal, and identify
the top layer (sheet 1) as the `superstrate' and the bottom layer
(sheet 2) as the thermally active sheet that dissipates heat to the
substrate. Hence, we write $D_{\text{top}}(\mathbf{q},\omega)=D_{\text{2D}}^{(1)}(\mathbf{q},\omega)$,
where 
\begin{equation}
D_{\text{2D}}^{(1)}(\mathbf{q},\omega)=[\rho_{1}\omega^{2}+i\rho_{1}\gamma_{1}\omega-\kappa_{1}q^{4}]^{-1}\label{Eq:Sheet1GreensFunction}
\end{equation}
is the retarded Green's function for the \emph{uncoupled} sheet 1
like in Eq.~(\ref{Eq:BareGreensFunc}) and $\rho_{1}$, $\gamma_{1}$
and $\kappa_{1}$ are the areal mass density, the frequency-dependent
flexural damping function and bending rigidity, respectively, for
sheet 1. We write the retarded Green's function for the bottom layer
(sheet 2), which is coupled to the top layer but uncoupled to the
substrate, as $\overline{D}_{\text{2D}}(\mathbf{q},\omega)=D_{\text{2D}}^{(2)}(\mathbf{q},\omega)$,
where
\begin{equation}
D_{\text{2D}}^{(2)}(\mathbf{q},\omega)=[\rho_{2}\omega^{2}+i\rho_{2}\gamma_{2}\omega-\kappa_{2}q^{4}-P_{2}]^{-1}\ ,\label{Eq:Sheet2GreensFunction}
\end{equation}
and 
\begin{equation}
P_{2}(\mathbf{q},\omega)=g_{2,1}[1-g_{2,1}D_{\text{2D}}^{(1)}(\mathbf{q},\omega)]^{-1}\label{Eq:SelfEnergy2}
\end{equation}
is the `self-energy' of sheet 2 from coupling to sheet 1 and corresponds
to the change in the flexural response of sheet 2 due to that coupling.
The terms $\rho_{2}$, $\gamma_{2}$ and $\kappa_{2}$ represent the
areal mass density, the frequency-dependent flexural damping function
and bending rigidity, respectively, for sheet 2 while $g_{2,1}$ is
the spring constant per unit area coupling sheets 1 and 2. Physically,
$D_{\text{2D}}^{(2)}(\mathbf{q},\omega)$ is the mechanical transfer
function that describes the flexural response of sheet 2 to the interfacial
stress exerted on its bottom surface by the substrate. 

We make the observation that in the  two-layer crystal, $D_{\text{2D}}^{(n)}(\mathbf{q},\omega)$
for $n=1,2$ describes the flexural response of sheet $n$ to the
interfacial stress exerted from the bottom and, for $n=2$,  includes
only the effects of mechanical coupling to the sheet immediately above
it via the self-energy term in Eq.~(\ref{Eq:SelfEnergy2}). Also,
the bottom layer (sheet 2) is the thermally active sheet from which
heat is dissipated to the substrate. This suggests that for the $N$-layer
crystal where the individual sheets are numbered from 1 (top) to $N$
(bottom) as shown in Fig.~\ref{Fig:ModelSchematics}(c), the flexural
response of the thermally active bottom layer (sheet $N$) is described
by the Green's function 
\begin{equation}
D_{\text{2D}}^{(N)}(\mathbf{q},\omega)=[\rho_{N}\omega^{2}+i\rho_{N}\gamma_{N}\omega-\kappa_{N}q^{4}-P_{N}]^{-1}\ ,\label{Eq:SheetNGreensFunction}
\end{equation}
where $P_{N}(\mathbf{q},\omega)=g_{N,N-1}[1-g_{N,N-1}D_{\text{2D}}^{(N-1)}(\mathbf{q},\omega)]^{-1}$
and $g_{N,N-1}$ is the spring constant per unit area coupling sheets
$N$ and $N-1$. In effect, $D_{\text{2D}}^{(N)}(\mathbf{q},\omega)$
depends on $D_{\text{2D}}^{(N-1)}(\mathbf{q},\omega)$ which in turn
depends on $D_{\text{2D}}^{(N-2)}(\mathbf{q},\omega)$ and so on.
Thus, we can recursively define $D_{\text{2D}}^{(n)}(\mathbf{q},\omega)$
for $n=N,N-1,\ldots,1$ with $D_{\text{2D}}^{(1)}(\mathbf{q},\omega)$
defined in Eq.~(\ref{Eq:Sheet1GreensFunction}) describing the flexural
response of sheet 1 at the top of the $N$-layer stack.

In the $N$-layer 2D crystal composed of identical sheets, we have
$\rho_{1}=\ldots=\rho_{N}=\rho$, $\kappa_{1}=\ldots=\kappa_{N}=\kappa$
and $\gamma_{1}=\ldots=\gamma_{N}=\gamma(\omega)$ while the interlayer
spring constant per unit area is given by $g_{N,N-1}=\ldots=g_{2,1}=g$,
and there are $N$ flexural phonon branches. Therefore, $D_{\text{2D}}^{(N)}(\mathbf{q},\omega)$,
the retarded Green's function for the flexural motion of the thermally
active bottom sheet, can be expressed as a weighted sum of the contributions
by the flexural phonon branches, \emph{i.e.}, 
\begin{equation}
D_{\text{2D}}^{(N)}(\mathbf{q},\omega)=\sum_{n=1}^{N}\frac{f_{n}}{\rho\omega^{2}+i\rho\gamma(\omega)\omega-\rho\omega_{n}^{(N)}(q)^{2}}\label{Eq:IdenticalSheetNGreensFunction}
\end{equation}
where $f_{1}=\frac{1}{N}$ and $f_{n}=\frac{1}{N}[1+\cos\frac{(n-1)\pi}{N}${]}
for $1<n\leq N$ are the weights. The details of the derivation of
Eq.~(\ref{Eq:IdenticalSheetNGreensFunction}) are given in the Appendix.
The term $\omega_{n}^{(N)}(q)$ describes the phonon dispersion for
the $n$-th branch and is given by

\begin{equation}
\omega_{n}^{(N)}(q)=\sqrt{\frac{\kappa}{\rho}q^{4}+(\Omega_{n}^{(N)})^{2}}\ ,\label{Eq:BranchPhononDispersion}
\end{equation}
where $\Omega_{n}^{(N)}=\sqrt{\frac{4g}{\rho}}\left|\sin\frac{(n-1)\pi}{2N}\right|$
is the $\Gamma$-point ($q=0$) phonon frequency for the $n$-th phonon
branch in the $N$-layer 2D crystal~\cite{Luo:PRB96_Theory,Zhao:NL13_Interlayer}.

\section{Results and discussion\label{Sec:Results-and-discussion}}

\subsection{Dependence of TBR on film thickness and temperature}

\begin{figure}
\includegraphics[width=8cm]{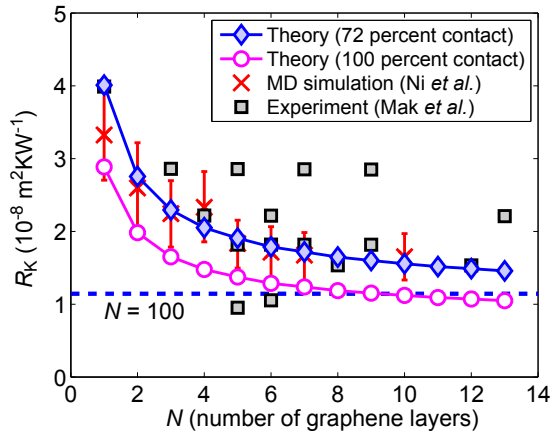}

\caption{Comparison of the theoretical room-temperature thermal boundary resistance
($R_{\text{K}}$) vs. number of graphene layers ($N$) (circles) with
data taken from experiments~\cite{Mak:APL10_Measurement} (squares)
and MD simulations~\cite{Ni:APL13_Few} (cross symbols). We also
plot the effective thermal boundary resistance ($R_{\text{K}}^{\prime}$)
corresponding to 72 percent contact (diamond symbols). The dashed
line ($1.14\times10^{-8}$ m$^{2}$KW$^{-1}$) corresponds to $R_{\text{K}}^{\prime}$
for $N=100$ and approximates the simulated $R_{\text{K}}^{\prime}$
for the graphite/SiO$_{2}$ interface.}

\label{Fig:TBRData}
\end{figure}

To evaluate the effectiveness of the theory, we compute $R_{\text{K}}$
for the graphene/SiO$_{2}$ interface for different $N$ using Eqs.~(\ref{Eq:ThermalBoundaryConductance})
and (\ref{Eq:IdenticalSheetNGreensFunction}), and plot the results
in Fig.~\ref{Fig:TBRData} alongside data from experiments~\cite{Mak:APL10_Measurement}
and MD simulations~\cite{Ni:APL13_Few} for comparison. The parameters
$c_{L}$, $c_{T}$, $\rho_{\text{sub}}$, $\rho$ and $\kappa$ and
the damping function $\gamma(\omega)$ for the flexural motion of
individual graphene sheets are taken from Ref.~\cite{Ong:PRB16_Theory}
while we set $g=1.095\times10^{20}$ Nm$^{-3}$ which we estimate
from the $\Gamma$-point interlayer breathing mode frequency ($\omega_{B}=11.2$
meV) for bilayer graphene~\cite{Yan:PRB08_Phonon} using the formula
$g=\frac{1}{2}\rho\omega_{B}^{2}$ derived from Eq.~(\ref{Eq:BranchPhononDispersion}).
In the $N\rightarrow\infty$ limit where we have graphite, the frequency
spacing between the flexural phonon dispersion curves vanishes and
this value of $g$ can be used to estimate $c_{z}$ the speed of sound
in the out-of-plane direction. Our estimate using the expression $c_{z}=a\sqrt{g/\rho}$,
where $a=0.335$ nm is the interlayer spacing in graphite, yields
$c_{z}=4021$ ms$^{-1}$, in excellent agreement with reported values~\cite{Prasher:PRB08_Thermal}. 

We also define the effective TBR ($R_{\text{K}}^{\prime}$), which
corresponds to the higher thermal resistance due to contact imperfections
at the interface~\cite{Huang:Carbon16_Improved,Liu:SP17_Thermal},
as $R_{\text{K}}^{\prime}=R_{\text{K}}/\chi$ where $0<\chi\leq1$
is the effective contact between graphene and the substrate, and plot
$R_{\text{K}}^{\prime}$ for $\chi=0.72$. Figure~\ref{Fig:TBRData}
shows $R_{\text{K}}$ exhibiting a thickness dependence qualitatively
similar to the MD simulation data~\cite{Ni:APL13_Few}: $R_{\text{K}}$
decreases as $N$ increases and converges asymptotically at large
$N$. Quantitative agreement between our theory and the MD simulation
data improves when we use $R_{\text{K}}^{\prime}$ instead of $R_{\text{K}}$,
indicating contact imperfection as a possible cause of discrepancy.
Other possible explanations for the discrepancy include: (1) the fact
that a continuum theory like ours cannot perfectly describe the dynamics
of an atomistic model especially outside the long-wavelength regime,
and (2) the variations in the effective elasticity parameters which
can lead to differences in the computed TBR values. For instance,
the Brenner-type interatomic potentials~\cite{Brenner:PRB90_Empirical}
used in Ref.~\cite{Ni:APL13_Few} have been shown~\cite{Lu:JPhysD09_Elastic}
to yield a smaller bending rigidity value which corresponds to a lower
numerical value for the TBC (or higher $R_{\text{K}}$) in our theory.
In addition, finite-size effects in MD simulations can be another
source of discrepancy.

The $R_{\text{K}}^{\prime}$ values also broadly agree with the experimental
data from Ref.~\cite{Mak:APL10_Measurement} although the spread
in the experimental values, possibly due to variation in the effective
contact of the interface ($\chi$) as well as other experimental factors,
makes it difficult for us to fit the data properly. Indeed, the simulated
$R_{\text{K}}$ curve corresponding to $\chi=1$ in Fig.~\ref{Fig:TBRData}
sets the lower bound for the range of TBR values extracted from Ref.~\cite{Mak:APL10_Measurement}
except at $N=5$ and $6$, suggesting that a significant percentage
of the interface may not be in perfect ($\chi=1$) contact for most
samples. Another possible cause of the poor disagreement is the corrugation
of the graphene/SiO$_{2}$ interface~\cite{Ishigami:NL07_Atomic}
which may introduce stochastic variation in the measured TBR not captured
by our theory. Modeling this corrugation effect on the TBR however
requires additional modification of our theory and a deeper understanding
of the connection between interfacial heat flow and contact mechanics,
which is beyond the scope of the current work although it is discussed
in Ref.~\cite{Persson:EPJE10_Heat}. We also obtain $R_{\text{K}}^{\prime}=1.14\times10^{-8}$
m$^{2}$KW$^{-1}$ for $N=100$ which approximates our projection
for the effective TBR of the room-temperature graphite/SiO$_{2}$
interface. This numerical value is comparable to the room-temperature
TBR value ($\sim10^{-8}$ m$^{2}$KW$^{-1}$) obtained by Schmidt
and co-workers~\cite{Schmidt:JAP10_Thermal} for the interface between
highly ordered pyrolytic graphite (HOPG) and Al film with a 5-nm Ti
adhesion layer although such a comparison between this TBR value and
our projected effective TBR for the graphite/SiO$_{2}$ interface
cannot be rigorously made as the substrate materials are different. 

\begin{figure}
\includegraphics[width=8cm]{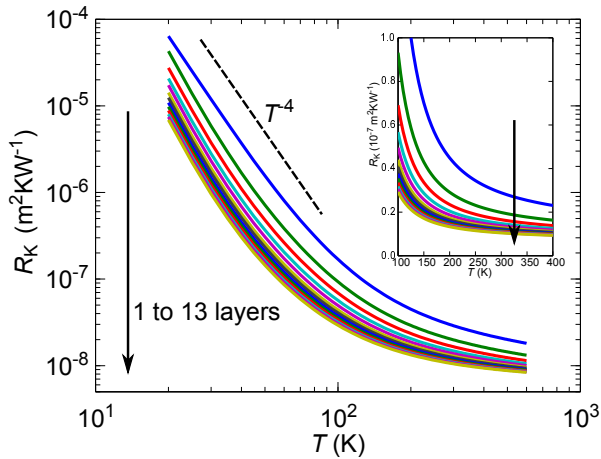}\caption{Thermal boundary resistance of the graphene/SiO$_{2}$ interface for
$N=1$ to $13$ and $T=20$ to $600$ K. The inset shows the same
data from $T=100$ to $400$ K with a linear scale. }

\label{Fig:TBC_temperature}
\end{figure}

Figure~\ref{Fig:TBC_temperature} also shows the TBR for $N=1$ to
$13$ and $T=20$ to $600$ K decreasing with temperature and scaling
approximately as $R_{\text{K}}\propto T^{-4}$ at low temperatures
instead of the typical $T^{-3}$ behavior~\cite{Swartz:RMP89_Thermal}
seen in nonlayered systems~\cite{Schleeh:NatMater15_Phonon,Zolotavin:ACSNano16_Plasmonic}.
The $R_{\text{K}}\propto T^{-4}$ behavior is also noted in Ref.~\cite{Ong:PRB16_Theory}
and is attributed to the $\omega^{3}$ scaling of the total interfacial
transmission $\Theta(\omega)$ at low frequencies. At low temperatures
($T\leq100$ K), the difference in $R_{\text{K}}$ for $N=1$ and
$N=13$ is significantly larger, by almost an order of magnitude,
than the difference at high temperatures ($T\geq300$ K).

\subsection{Dependence of the interfacial transmission spectrum on film thickness}

\begin{figure}
\includegraphics[width=8.5cm]{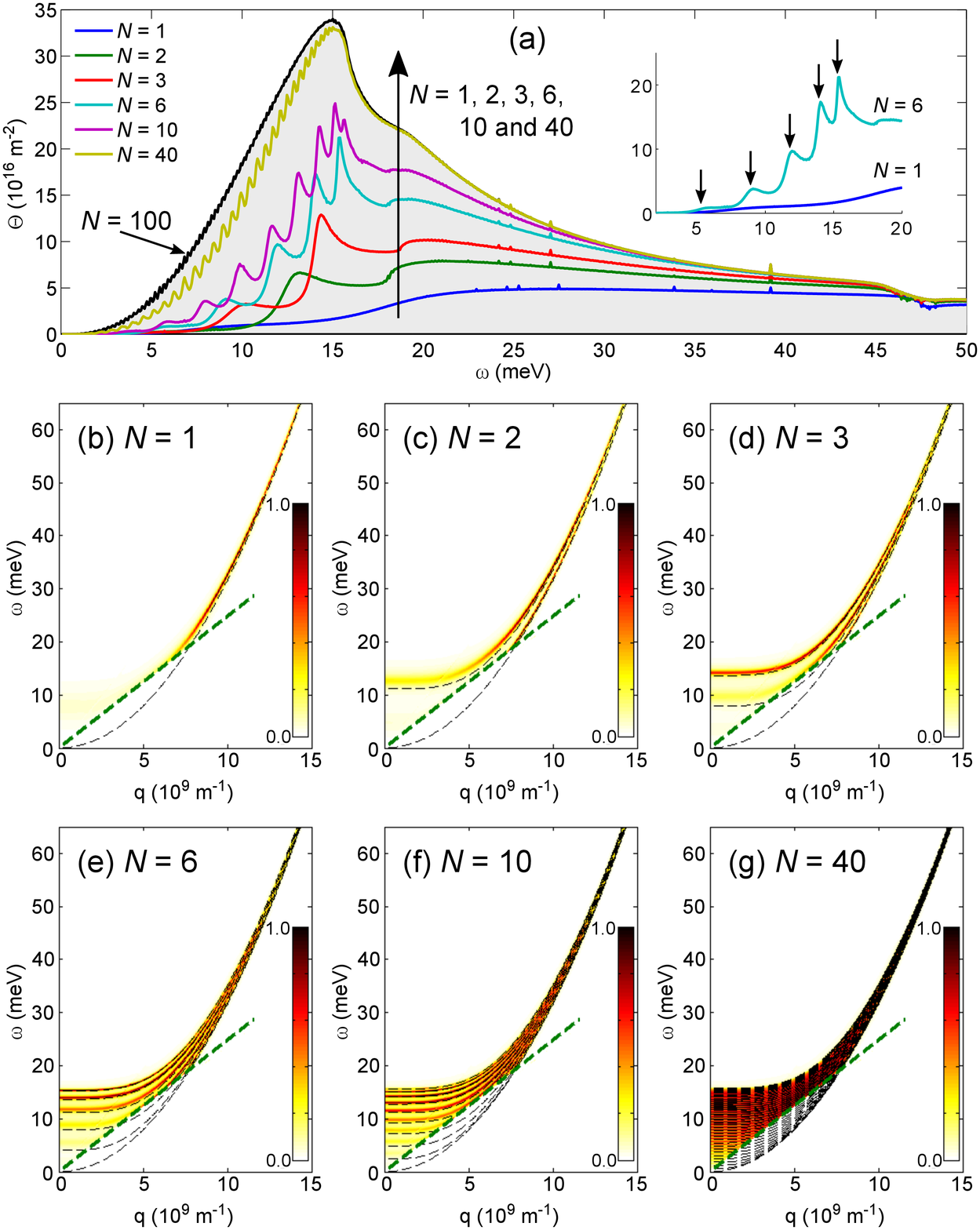}

\caption{(a) Overall transmission per unit area $\Theta(\omega)$ for $N=1$,
$2$, $3$, $6$, $10$, $40$ and $100$, with the $N=100$ spectrum
(shaded gray) approximating the graphite/SiO$_{2}$ spectrum. The
inset compares the low-frequency spectra for $N=1$ and $N=6$ with
peak positions in the latter indicated by arrows. We also plot the
transmission coefficient spectrum $\Xi(\mathbf{q},\omega)$ for (b)
$1$-, (c) $2$-, (d) $3$-, (e) $6$-, (f) $10$-, and (g) $40$-layer
graphene. The thin black dashed lines correspond to the flexural phonon
branch dispersion given by Eq.~(\ref{Eq:BranchPhononDispersion}),
while the thicker green dashed lines correspond to the bulk transverse
acoustic phonon dispersion for the substrate ($\omega=c_{T}q$). }
\label{Fig:TransmissionSpectra}
\end{figure}

The TBR reduction in thicker samples shown in Fig.~\ref{Fig:TBRData}
can be interpreted in terms of changes in the overall phonon transmission
between graphene and the SiO$_{2}$ substrate. To quantify the overall
phonon transmission, we define the frequency-dependent transmission
function per unit area~\cite{Ong:PRB16_Theory} 
\begin{equation}
\Theta(\omega)=\frac{1}{(2\pi)^{2}}\int d^{2}q\Xi(\mathbf{q},\omega)\ .\label{Eq:TransmissionPerArea}
\end{equation}
Figure~\ref{Fig:TransmissionSpectra}(a) shows $\Theta(\omega)$
for $N=1$, $2$, $3$, $6$, $10$, $40$ and $100$, with the $N=100$
spectrum approximating the graphite/SiO$_{2}$ interface. We find
that as $N$ increases, the additional layers result in substantial
enhancement of $\Theta(\omega)$ especially at lower frequencies.
Hence, we attribute the decrease in TBR for few-layer graphene to
the progressively enhanced cross-plane transmission of low-frequency
modes as $N$ increases. In addition, we also observe a number of
peaks in $\Theta(\omega)$ which increases as $N$ gets larger. This
can be more clearly seen in the inset of Fig.~\ref{Fig:TransmissionSpectra}(a)
where the smooth $N=1$ spectrum is contrasted with the $N=6$ spectrum
with  five peaks, of which the higher frequency ones are sharper and
more closely spaced. In general, there are $N-1$ peaks in the transmission
spectrum of $N$-layer graphene although the peaks gradually merge
together when $N$ is large as is evident from the $N=40$ and $100$
spectra which are practically identical for $\omega>15$ meV.

The origin of the enhanced phonon transmission and the transmission
peaks can be connected to the flexural phonon dispersion in $N$-layer
graphene by analyzing the transmission coefficient function $\Xi(\mathbf{q},\omega)$
which we plot for different values of $N$ in Fig.~\ref{Fig:TransmissionSpectra}(b)-\ref{Fig:TransmissionSpectra}(g).
In  one-layer graphene {[}Fig.~\ref{Fig:TransmissionSpectra}(b){]},
the low-$q$ part of the transmission coefficient spectrum corresponding
to the only flexural phonon branch is insignificant because the associated
phonon frequencies are less than the substrate bulk transverse acoustic
frequencies, i.e. $\omega<c_{T}q$, which results in weak coupling
between the low-frequency flexural phonon modes and the substrate
as pointed out in Ref.~\cite{Ong:PRB16_Theory} In  two-layer graphene
{[}Fig.~\ref{Fig:TransmissionSpectra}(c){]}, the lower phonon branch
does not contribute significantly to interfacial transmission like
in  one-layer graphene since $\omega<c_{T}q$ but the upper phonon
branch contribution is more pronounced in the low-$q$ part of the
transmission spectrum as the associated upper branch modes satisfy
the condition $\omega>c_{T}q$. The contribution from the higher phonon
branches becomes more substantial for three, six and ten-layer graphene
{[}Figs.~\ref{Fig:TransmissionSpectra}(d)-\ref{Fig:TransmissionSpectra}(f){]}.
In particular, the higher phonon branches show sharper transmission
peaks, indicating more efficient interfacial transmission. Therefore,
we attribute the enhanced cross-plane phonon transmission and hence
the lower TBR of a thicker graphene film to the interfacial transmission
contribution from its additional flexural phonon branches, which becomes
more pronounced for low frequencies and produces the greater low-temperature
$R_{K}$ change with respect to $N$ seen in Fig.~\ref{Fig:TBC_temperature}.
Also, the transmission peaks in Fig.~\ref{Fig:TransmissionSpectra}(a)
are due to the long wavelength ($q\rightarrow0$) modes of the $N-1$
higher flexural branches in $N$-layer graphene. 

However, when $N$ is large {[}e.g., $N=40$ in Fig.~\ref{Fig:TransmissionSpectra}(g){]},
the contribution to interfacial transmission becomes much less dependent
on $N$. This can be explained by noting that the transmission coefficient
function in Eq.~(\ref{Eq:TransmissionFunction}) can be expressed
as the sum of the transmission by each phonon branch, \emph{i.e.},
\begin{equation}
\Xi(\mathbf{q},\omega)=\sum_{n=1}^{N}\frac{4K^{2}\text{Im}D_{\text{sub}}(\mathbf{q},\omega)\text{Im}D_{n}(\mathbf{q},\omega)}{|1-K[D_{\text{sub}}(\mathbf{q},\omega)+D_{\text{2D}}(\mathbf{q},\omega)]|^{2}}\label{Eq:TransmissionSum}
\end{equation}
where $D_{n}(\mathbf{q},\omega)$ is the summand in Eq.~(\ref{Eq:IdenticalSheetNGreensFunction})
and is proportional to the weight $f_{n}$ which scales as $N^{-1}$.
Therefore, at large $N$, the transmission contribution in Eq.~(\ref{Eq:TransmissionSum})
by each closely spaced phonon branch scales as $N^{-1}$ and thus,
the total transmission and the TBR converge asymptotically with respect
to $N$ because the additional transmission contribution from having
more phonon branches is canceled out by the diminishing contribution
of each branch. This weak $N$ dependence is also evident from the
very similar transmission spectra for $N=40$ and $100$ in Fig.~\ref{Fig:TransmissionSpectra}(a).

\section{Summary and conclusions}

In summary, we have generalized the theory of heat dissipation in
Ref.~\cite{Ong:PRB16_Theory} to multilayer 2D crystals and used
it to evaluate the TBR of the graphene/SiO$_{2}$ interface for different
layer numbers and temperatures. The key idea in the generalization
is to treat the top $N-1$ layers in an $N$-layer stack as a superstrate.
The numerical results show that the TBR decreases with $N$, consistent
with what has been observed in MD simulations and experiments, and
converges asymptotically to a constant corresponding to the graphite/SiO$_{2}$
TBR. Our theory explains (1) the TBR reduction with respect to layer
number in terms of the enhanced interfacial transmission contribution
by the additional flexural phonon branches and (2) its asymptotic
convergence, and predicts a low-temperature $R_{\text{K}}\propto T^{-4}$
behavior. It can also opens up new possibilities of heat dissipation
control through modification of the layered structure.
\begin{acknowledgments}
This work was supported in part by a grant from the Science and Engineering
Research Council (152-70-00017) and financial support from the Agency
for Science, Technology and Research (A{*}STAR), Singapore.
\end{acknowledgments}

\appendix*

\section{Derivation of $D_{\text{2D}}^{(N)}(\mathbf{q},\omega)$\label{Append}}

Here, we derive $D_{\text{2D}}^{(N)}(\mathbf{q},\omega)$ in Eq.~(\ref{Eq:IdenticalSheetNGreensFunction}).
The equations of motion for layers $1$ to $N$ in the $N$-layer
system can be written as \begin{subequations} 
\begin{equation}
-(\rho\omega^{2}+i\rho\gamma\omega-\kappa q^{4})u_{1}=-g(u_{1}-u_{2})\ ,
\end{equation}
\begin{equation}
-(\rho\omega^{2}+i\rho\gamma\omega-\kappa q^{4})u_{N}=-g(u_{N}-u_{N-1})
\end{equation}
\begin{equation}
-(\rho\omega^{2}+i\rho\gamma\omega-\kappa q^{4})u_{j}=-g(2u_{j}-u_{j-1}-u_{j+1})\ ,
\end{equation}
\label{Eq:EqnMotionNLayers}\end{subequations} where $j=2,\ldots,N-1$
and $u_{l}=u_{l}(\mathbf{q},\omega)$ is the flexural displacement
of the $l$-th layer. We can express Eq.~(\ref{Eq:EqnMotionNLayers})
more compactly as an eigenvalue equation
\begin{equation}
\sum_{j=1}^{N}(V_{ij}-z\delta_{ij})u_{j}=0\ ,\label{Eq:EigenvalueForm}
\end{equation}
where $z(\mathbf{q},\omega)=\rho\omega^{2}+i\rho\gamma(\omega)\omega-\kappa q^{4}$,
$V_{ij}=-g(\delta_{i,j-1}+\delta_{i,j+1})$ for $i\neq j$ and $V_{ii}=\sum_{j=1}^{N}(\delta_{ij}-1)V_{ij}$
for $i=1,\ldots,N$. The matrix $V$ is an $N\times N$ tridiagonal
matrix in which the off-diagonal elements have a value of $-g$ and
the diagonal elements are equal to the negative sum of the off-diagonal
values in the row. Therefore, the $n$-th eigenvalue of $V$ in Eq.~(\ref{Eq:EigenvalueForm})
is given by~\cite{Luo:PRB96_Theory} $z_{n}=4g\sin^{2}\left[\frac{(n-1)\pi}{2N}\right]$
while the flexural displacement of the $j$-th layer for the corresponding
$n$-th normalized eigenvector is 
\begin{equation}
u_{j}^{n}=\sqrt{\frac{2}{N}}\cos\left[\frac{(n-1)(2j-1)\pi}{2N}\right]\label{Eq:Eigenvector_n}
\end{equation}
for $n=2,\ldots,N$ and $u_{j}^{n}=1/\sqrt{N}$ for $n=1$. The $N\times N$
matrix describing the Green's function of the system can be written
as a resolvent, \emph{i.e.}, $G_{ij}(z)=\left(\frac{1}{z-V}\right)_{ij}=\sum_{n=1}^{N}(u_{i}^{n})^{\dagger}u_{j}^{n}(z-z_{n})^{-1}$.
Thus, the flexural response of the bottommost layer ($j=N$) is 
\begin{equation}
D_{\text{2D}}^{(N)}(\mathbf{q},\omega)=G_{NN}(z(\mathbf{q},\omega))=\sum_{n=1}^{N}\frac{f_{n}}{z(\mathbf{q},\omega)-z_{n}}\ ,\label{Eq:FlexuralResponseDefn}
\end{equation}
where $f_{1}=\frac{1}{N}$ and $f_{n}=\frac{1}{N}[1+\cos\frac{(n-1)\pi}{N}]$
for $1<n\leq N$. In the large $N$ limit, we can write Eq.~(\ref{Eq:FlexuralResponseDefn})
as an integral, \emph{i.e.}, 
\begin{equation}
\lim_{N\rightarrow\infty}D_{\text{2D}}^{(N)}(\mathbf{q},\omega)=\frac{1}{\pi}\int_{0}^{\pi}d\theta\frac{1+\cos\theta}{z(\mathbf{q},\omega)-4g\sin^{2}(\frac{1}{2}\theta)}\ .\label{Eq:LargeNLimit}
\end{equation}

\bibliographystyle{apsrev4-1}
\bibliography{references}

\end{document}